# Ambi-chiral anomalous Hall effect in magnetically doped topological insulators

Chang Liu[1,2†], Yunyi Zang[1,2†], Yan Gong[1], Ke He[1,2,3*], Xucun Ma[1,3], Qi-Kun Xue[1,2,3], Yayu Wang[1,3*]

[1]*State Key Laboratory of Low Dimensional Quantum Physics, Department of Physics, Tsinghua University, Beijing 100084, P. R. China*

[2]*Beijing Academy of Quantum Information Sciences, Beijing 100193, P. R. China*

[3]*Frontier Science Center for Quantum Information, Beijing 100084, P. R. China*

[†] *These authors contributed equally to this work.*

[*] Emails: kehe@tsinghua.edu.cn; yayuwang@tsinghua.edu.cn

**Abstract:** The chirality associated with broken time reversal symmetry in magnetically doped topological insulators have important implications to the quantum transport phenomena. Here we report the anomalous Hall effect studies in Mn- and Cr-doped $Bi_2Te_3$ topological insulators with varied thickness and doping content. By tracing the chirality of the Hall loops, we find that the Mn-type anomalous Hall effect with clockwise chirality is strengthened by the reduction of film thickness, which is opposite to that of the Cr-type anomalous Hall effect with counterclockwise chirality. We provide a phenomenological model to explain the evolution of magnetic order and anomalous Hall effect chirality in magnetically doped topological insulators.

**Introduction**

Chirality is defined as an asymmetric property that is used to describe a system that cannot be distinguished from its mirror image. It plays a crucial role in a wide range of systems, ranging from fundamental particles in high energy physics, to grand-scale objects like spiral galaxies in astrophysics. In condensed matter physics, chiral phenomena not only manifest itself in material structure but also in emergent electromagnetic responses [1-4]. Magnetic topological insulators (TIs) represent versatile platforms for exploring the exotic quantum phenomena with chiral nature. For example, ferromagnetic (FM) order breaks the time-reversal symmetry (TRS) of TIs and endows the Dirac fermions with finite mass through the exchange coupling [5-12]. The helicity of the topological surface states is lifted and the resulting nonzero Berry curvature leads to pronounced anomalous Hall effect (AHE) that exhibits hysteresis with a specific chirality. In the two-dimensional (2D) limit when Fermi level is tuned to the band gap, the one-dimensional (1D) helical channels of TIs is split into chiral edge state that contributes to the quantized AHE [13-15]. The realization of quantized AHE undoubtedly demonstrates the critical roles of Berry curvature on the electronic properties of TIs, and provides a new building block for realizing other exotic topological quantum phenomena such as magnetic monopoles and chiral Majorana fermions [16-24].

Despite considerable efforts on magnetic TIs, particularly in pursuing the quantized AHE, there are still intriguing open questions. In previous studies, the AHE behaviors in magnetic TIs can be divided into two different groups. In the first group, exemplified by the Cr- and V-doped $Bi_2Te_3$ family TI films, the magnetic field dependent Hall traces display hysteresis loops with counterclockwise chirality [13-15,25-27], as schematically shown on the left of Figure 1(a). Here, $\rho_{yx}^+$ and $\rho_{yx}^-$ denote the Hall resistivity $\rho_{yx}$ profiles when magnetic field is swept back to zero from positive and negative polarizations respectively. The violet arrows represent the magnetization at corresponding magnetic field. At positive magnetization, the AH resistivity $\rho_{yx}^0$ (defined as $\rho_{yx}^+$ at zero magnetic field) has a positive value. For the quantized AHE such as in the Cr-doped $(Bi,Sb)_2Te_3$ TI system, $\rho_{yx}^0$ reaches $h/e^2$, where $h$ denotes the Plank constant and $e$ is the electron charge, corresponding to the Chern number $C = +1$ [13-15]. In the second group, represented by Mn-doped $Bi_2Te_3$, the Hall traces display hysteresis loop with clockwise chirality and the sign of $\rho_{yx}^0$ is negative [28-30]. Thus, for its quantized AHE version, the Chern number is $C = -1$ and $\rho_{yx}^0$ is quantized at $-h/e^2$, which has recently been realized in MnBi$_2$Te$_4$ intrinsic antiferromagnetic TI with odd number of layers [31]. The schematic picture of the AHE with clockwise chirality is displayed on the right of Figure 1(a).

Accompanied by the rise of chiral spintronics in topological matter [4], the chirality issue of the AHE has attracted more attention in recent years. A variety of exotic topological quantum phenomenon are found to be associated with chirality [18,28,32-37]. For example, theoretical calculations propose that the opposite chirality of the AHE in Mn- and Cr-doped Bi$_2$Te$_3$ can be utilized to realize the long-sought quantized topological magnetoelectric effect when they are constructed into heterostructures [18]. Very recently, transport studies in pulsed magnetic field demonstrates that the Chern insulator phase in Mn-based magnetic TI with clockwise chirality and $C = -1$ evolves into a helical phase with $C = 0$ [31,38,39]. In contrast, the Chern insulator phase in Cr-based magnetic TI with counterclockwise chirality and $C = +1$ becomes more stable in the high field limit [13,32]. These results provide more motivations for understanding and controlling the chirality of AHE in magnetic TIs.

Although there have been several experiments studying the AHE in Mn- and Cr-doped TIs [13-15,25-29,40,41], for a certain magnetic TI thin film, the chirality of AHE is usually fixed. It is difficult to switch between opposite chirality in one sample. Moreover, the different sample fabrication methods [42,43], selections of dopants and parent TI compounds [43-49], effects of interface [33,34], and even dimensionality [28,29,40,41], make it elusive to figure out how the AHE with opposite chirality evolves under different parameters. Therefore, it is highly desired to have an accurate control of parameters in one system to study the chirality of the AHE.

In this work, we report systematic transport studies of Cr- and Mn-doped (Bi$_{0.9}$Mn$_x$Cr$_{0.1-x}$)$_2$Te$_3$ magnetic TIs with various thickness $d$ and doping level $x$ at different temperature $T$ and gate voltage $V_g$. We find that the Mn-type AHE with clockwise chirality is strengthened as $d$ is reduced from 8 quintuple-layer (QL) to 6 QL, whereas the Cr-type AHE with counterclockwise chirality exhibits totally opposite behavior. The different response of AHE to film thickness are universally present in the $x$, $T$ and $V_g$ dependent measurements, consistent with the distinct mechanisms for FM order found in Mn- and Cr-doped TIs. We propose a simple model to explain the observed ambi-chiral AHE in magnetically doped TIs.

## Methods

The ten magnetically doped TI samples studied in this work are grown by molecular beam epitaxy on insulating SrTiO$_3$(111) substrates following the similar method reported previously [13,28,50]. Mn and Cr are uniformly deposited during the sample growth. The SrTiO$_3$ substrate also acts as a back gate due to its large dielectric constant at low $T$. Figure 1(b) shows the crystal structure and the schematic cross-section of a (Bi$_{0.9}$Mn$_x$Cr$_{0.1-x}$)$_2$Te$_3$ TI. The black and red arrows mark the Mn and Cr magnetic moments respectively. Two film thicknesses $d$ = 8 QL and 6 QL are chosen because they are small enough to ensure sufficient gating ability and large enough to avoid the effect of surface hybridization on the magnetic properties [28,30,51].

For transport measurement, standard four-probe AC lock-in method is carried out after the samples are manually scratched into Hall bar structures. Indium pieces are pressed directly on the top surface of the samples as electrical contacts. The magnetic field is applied perpendicular to the sample plane, and the Hall signals are antisymmetrized with respect to magnetic field to correct the errors caused by geometric misalignments. An excitation current of 0.2 μA is applied by a Keithley 6221 current source, and a $V_g$ is applied by a Keithley 2400 source meter.

## Experimental results

We first study the magnetic field dependent $\rho_{yx}$ at different $T$s for five 8-QL (Bi$_{0.9}$Mn$_x$Cr$_{0.1-x}$)$_2$Te$_3$ samples, as shown in Figure 2(a). With the decrease of $T$, $\rho_{yx}$ exhibits a hysteresis loop as magnetic field is swept back and forth, indicating the formation of FM order. The blue and red arrows denote the opposite chirality of the Hall traces. In the purely Cr-doped Bi$_2$Te$_3$ sample ($x$ = 0), $\rho_{yx}^0$ reaches ~1.5 kΩ at $T$ = 1.6 K. With the increase of $x$, $\rho_{yx}^0$ progressively decreases as more Cr dopants are substituted by Mn. At $x$ = 0.08, $\rho_{yx}^0$ is suppressed to a vanishingly small value. As $x$ is further increased to 0.10, the sample becomes purely Mn-doped, in which $\rho_{yx}^0$ turns to negative and the chirality of the hysteresis loop becomes clockwise.

In the next, we study $\rho_{yx}$ in another set of samples but with $d$ = 6 QL, where the contribution from bulk state is reduced. The overall behaviors are similar to that of the 8-QL samples, but there is an obvious difference in the $x$ value when the $\rho_{yx}^0$ sign change occurs. As indicated by the data set enclosed by the red frames, it is reduced from $x = 0.08$ in the 8-QL samples to $x = 0.07$ in the 6-QL samples. To better visualize the thickness dependent chirality change, we plot the difference between $\rho_{yx}^+$ and $\rho_{yx}^-$ (defined as $\triangle\rho_{yx} = \rho_{yx}^+ - \rho_{yx}^-$) as a function of magnetic field for different $x$ and $d$, as shown in Figures 2(c) and 2(d). The peak and valley in $\triangle\rho_{yx}$ near zero field directly reflects the chirality of AHE. Clearly, in the thinner samples with $d = 6$ QL, less Mn is required to tune the chirality from counterclockwise to clockwise. The effect of film thickness on the AHE is also reflected by the value of $\rho_{yx}^0$. In the purely Cr-doped sample, with the reduction of $d$, $|\rho_{yx}^0|$ decreases from ~0.85 k$\Omega$ to ~0.35 k$\Omega$ at $T = 1.6$ K. In contrast, $|\rho_{yx}^0|$ increases from ~0.15 k$\Omega$ to ~0.25 k$\Omega$ in the purely Mn-doped sample. The above results suggest that the Mn-type clockwise AHE is strengthened in thinner samples, whereas the Cr-type AHE with counterclockwise chirality is preferred in thicker ones.

Because tuning the doping contents can only tune the AHE in a discrete way, to gain more detailed knowledge about the chirality change, we explore the magnetic field dependent $\rho_{yx}$ at varied $V_g$s in these samples. Figures 3(a) and 3(b) display the main results for the two critically doped samples ($x = 0.08$ for 8-QL and $x = 0.07$ for 6-QL) that are close to the boundary of the chirality change. Interestingly, the chirality in the two samples exhibit quite different evolutions with the variation of $V_g$. For the 8-QL sample, the chirality remains counterclockwise for the entire $V_g$ range, suggesting the Cr-related AHE dominates the transport. In contrast, for the 6-QL sample, increasing $V_g$ dramatically changes the shape of $\rho_{yx}$ loop and even cause a change of chirality. At $V_g = -100$ V, the hysteresis is clockwise, and the sign of $\rho_{yx}^0$ is negative. As $V_g$ increases to -25 V, the two $\rho_{yx}$ curves from opposite field sweeps touch together at zero field, forming a hysteresis with ill-defined chirality. As $V_g$ is further increased to 100 V, the two $\rho_{yx}$ curves form a hysteresis loop again but now with counterclockwise chirality and positive $\rho_{yx}^0$. In the insets of Figure 3(b), we plot the zoomed-in data in the field range of 0.3 T, which clearly shows the change of chirality with $V_g$.

To visualize the $V_g$ dependent ambi-chiral AHE more clearly, we display the magnetic field dependent $\triangle\rho_{yx}$ for varied $V_g$, as shown in Figures 4(a) and 4(b). In the 8-QL sample, a positive peak near zero field is universally present throughout the $V_g$ range, in accord with the positive sign of $\rho_{yx}^0$ ($\rho_{yx}^+ > 0$ at zero field). However, in the 6-QL sample, the crossing of $\rho_{yx}^+$ and $\rho_{yx}^-$ curves give rise to a more complex evolution in $\triangle\rho_{yx}$, as displayed in Figure 4(b). With the decrease of $V_g$, the peak value at zero field shifts downward, accompanied by the sign change from positive to negative. Eventually, $\triangle\rho_{yx}$ becomes an overall negative curve characterized by two negative dips near zero field. The above $V_g$ dependent chirality change is highly reminiscent of the doping dependent chirality change as shown in Figures 2(c) and 2(d).

To further illustrate the distinct dimensionality dependence of Mn- and Cr-type AHE, in Figures 4(c) and 4(d), we display the color maps of $\triangle\rho_{yx}$ as functions of magnetic field and $V_g$ for samples with different $x$ values. In each figure, $d$ is fixed and $x$ increases from panel (i) to (iii). The dotted data represent the position where $\rho_{yx}^+$ and $\rho_{yx}^-$ crosses, and the red and blue colors represent the area with $\triangle\rho_{yx} > 0$ and $\triangle\rho_{yx} < 0$. In the purely Mn-doped $Bi_2Te_3$, $\rho_{yx}^+$ is always smaller than $\rho_{yx}^-$ in the hysteretic regime, thus $\triangle\rho_{yx}$ is negative near zero field. Whereas in the purely Cr-doped sample, $\triangle\rho_{yx}$ has a positive value. Therefore, we can use the sign of $\triangle\rho_{yx}$ to characterize the Mn- and Cr-type AHE. Clearly, at $d$ = 8 QL, the Mn-type AHE is not favored until $x$ is increased to 0.08 and when negative $V_g$ is applied. However, in the 6-QL samples, a smaller Mn doping level ($x$ = 0.05) is sufficient to induce similar trend in $\triangle\rho_{yx}$, as also shown by the growing area with $\triangle\rho_{yx} > 0$ (red). Again, it suggests that the Mn-type AHE is strengthened in thinner samples, opposite to that of the Cr-type AHE.

## Discussions and conclusions

Previous theories have proposed that the topologically protected surface states in TIs could be more susceptible to magnetic order than the insulating bulk states, thus is anticipated to give rise to distinct magnetic ordering phenomenon in Mn-doped $Bi_2Se_3$ TIs [52]. In fact, there has been several transport and spectroscopy experiments suggesting that the FM order in Mn-doped TIs is mediated by the surface states [30,44,46-48], while it is the bulk that mainly contributes to FM order in Cr-doped system [26,42,43,53]. The opposite dimensionality dependence of the Mn- and Cr-related AHE observed here suggests a scenario of dopant-selective surface and bulk magnetic order in Mn- and Cr-doped $Bi_2Te_3$, consistent with previous reports. In Figures 5(a) and 5(b), we summarize the $T$ dependent $\rho_{yx}^0$ for all ten samples, and the two insets show the zoomed-in data for the two critical samples. Clearly, not only the value of $|\rho_{yx}^0|$ but also the onset temperature ($T_{AH}$) responses oppositely to thickness in Mn- and Cr-doped $Bi_2Te_3$. As $d$ is reduced from 8 QL to 6 QL, $T_{AH}$ of the Mn-Bi$_2$Te$_3$ sample remains almost unchanged (red arrows). However, $T_{AH}$ of the Cr-Bi$_2$Te$_3$ sample decreases from 26 K to 14 K (black arrows). Such effect is more clearly visualized in the color plot of $\rho_{yx}^0$ in Figures 5(c) and 5(d). These results demonstrate that reducing $d$ from 8 QL to 6 QL have more dramatic effect on the Cr-type AHE than the Mn-type one. According to previous measurements [51], the penetration depth of the topological surface state is about 2 QL, thus reducing $d$ from 8 QL to 6 QL mainly suppresses the bulk volume while keeps the surface states less-affected. These results strongly indicate the dimensionality dependent chirality change in our samples is related to the different surface and bulk FM order in Mn- and Cr-doped TIs [26,28,30,42-44,46-48,53].

Finally, we discuss about the possible factors that lead to the opposite chirality in Mn- and Cr-doped TIs. Theoretically, the AHE in magnetic TIs is induced through the exchange coupling $JM_zS_z$ between local moments $M_z$ and electron spins $S_z$, where $J$ denotes the coupling constant [5,18,30]. For fixed direction of $M_z$, opposite $J$ could endow Dirac fermions with opposite mass, thus leading to opposite sign and chirality of the AHE. However, the microscopic mechanism of the sign of $J$ still remains unclear. Here, we propose a phenomenological picture to address the chirality issue in terms of the $d$-orbit configurations of Mn and Cr dopants. According to previous studies, the spin states of Mn and Cr dopants in TIs are $S = 5/2$ and $3/2$ respectively [54,55]. Therefore, in Mn-based TIs, each of the five $3d$-orbits of Mn$^{2+}$ is singly occupied. When the itinerant electron spins couple with Mn$^{2+}$, only electron with opposite spin is allowed to virtually hop to the unoccupied orbits. However, in Cr-based TIs, each Cr$^{3+}$ has three singly occupied and two empty orbits. In this case, it is energetically more favorable for electrons with the same spin to hop to the empty orbits. Consequently, the exchange coupling and the resulting Dirac fermion mass in the two systems are expected to have opposite sign, which in turn leads to opposite chirality in their AHE. The above physical pictures are schematically illustrated in Figures 5(e) and 5(f) respectively.

In summary, we perform systematic studies of the AHE in Mn- and Cr-codoped Bi$_2$Te$_3$ TI thin films. By varying the Cr/Mn doping levels and film thickness, we can actively control the chirality of the AHE. These results can be interpreted by the different mechanisms for FM order and AHE chirality in Mn- and Cr-doped TIs. Our experiments shed new light on manipulating the interplay between FM order and band topology in magnetic TIs.

**FIGURE CAPTIONS**

**Figure 1** (a) Two types of AHE with counterclockwise and clockwise Hall traces. $\rho_{yx}^+$ and $\rho_{yx}^-$ denote the Hall resistivity acquired as magnetic field swept back from positive and negative polarization. The vertical violet arrows represent the magnetization at according magnetic field. (b) Schematic illustrations of crystal structure and cross-section of a $(Bi_{0.9}Mn_xCr_{0.1-x})_2Te_3$ device. The red and black arrows denote the Cr and Mn magnetic moments.

**Figure 2** (a)-(b) Magnetic field dependent $\rho_{yx}$ for ten $(Bi_{0.9}Mn_xCr_{0.1-x})_2Te_3$ magnetic TIs with different thickness at varied temperatures. The blue and red color represent the different chirality of Hall traces. (c)-(d) Magnetic field dependent $\triangle\rho_{yx}$ for samples with different $d$ and $x$.

**Figure 3** Magnetic field dependent $\rho_{yx}$ for an 8-QL sample with $x = 0.08$ (a), and a 6-QL sample with $x = 0.07$ (b) for varied $V_g$s. The blue and red color represent the data obtained from different field sweep direction, as labeled by $\rho_{yx}^+$ and $\rho_{yx}^-$ respectively. The insets present the zoomed-in data at low magnetic field range of 0.3 T. All the data are acquired at $T = 1.6$ K.

**Figure 4** Discrepancies between $\rho_{yx}^+$ and $\rho_{yx}^-$ for an 8-QL sample with $x = 0.08$ (a), and a 6-QL sample with $x = 0.07$ (b). (c) Color plots of $\triangle\rho_{yx}$ as a function of magnetic field and $V_g$ for three 8-QL samples with $x = 0.06$ (i), $x = 0.08$ (ii), and $x = 0.1$ (iii). The red and blue regimes represent the areas for Cr- and Mn-type AHE. (d) Color plots of $\triangle\rho_{yx}$ for three 6-QL samples with $x = 0.05$ (i), $x = 0.07$ (ii), and $x = 0.8$ (iii). The dotted data represent the position where Hall traces of opposite field sweep cross.

**Figure 5** (a)-(b) Temperature dependent $\rho_{yx}^0$ for ten samples with different thickness and doping level. The insets display the zoomed-in data for the two critical samples with $x = 0.08$ for the 8-QL sample and $x = 0.07$ for the 6-QL sample. (c)-(d) Color plots of $\rho_{yx}^0$ as functions of doping level and temperature. The red and blue colors represent the regimes for $\rho_{yx}^0 > 0$ and $\rho_{yx}^0 < 0$, respectively. The magenta dots denote the onset temperatures of AHE in purely Mn- and Cr-doped $Bi_2Te_3$ samples. (e)-(f) Schematic pictures of the Dirac gap opening process in Mn- and Cr-based TI systems. The blue and red arrows represent the spin states for localized *d*-electrons and itinerant electron respectively.

**Acknowledgements:** This work is supported by the Basic Science Center Project of NSFC (grant No. 51788104) and National Key R&D Program of China (grants No. 2018YFA0307100 and No. 2018YFA0305603). This work is supported in part by the Beijing Advanced Innovation Center for Future Chip (ICFC). Chang Liu is grateful to Peizhe Tang from Beihang University for useful discussions.

**Reference**

[1] X. Z. Yu, Y. Onose, N. Kanazawa, J. H. Park, J. H. Han, Y. Matsui, N. Nagaosa, and Y. Tokura, Nature **465**, 901 (2010).

[2] S. Muhlbauer, B. Binz, F. Jonietz, C. Pfleiderer, A. Rosch, A. Neubauer, R. Georgii, and P. Boni, Science **323**, 915 (2009).

[3] C. Train, R. Gheorghe, V. Krstic, L. M. Chamoreau, N. S. Ovanesyan, G. L. Rikken, M. Gruselle, and M. Verdaguer, Nat. Mater. **7**, 729 (2008).

[4] S. H. Yang, R. Naaman, Y. Paltiel, and S. S. P. Parkin, Nat. Rev. Phys. **3**, 328 (2021).

[5] Y. Tokura, K. Yasuda, and A. Tsukazaki, Nat. Rev. Phys. **1**, 126 (2019).

[6] X. L. Qi and S. C. Zhang, Rev. Mod. Phys. **83**, 1057 (2011).

[7] M. Z. Hasan and C. L. Kane, Rev. Mod. Phys. **82**, 3045 (2010).

[8] X. L. Qi, T. L. Hughes, and S. C. Zhang, Phys. Rev. B **78**, 195424 (2008).

[9] R. Yu, W. Zhang, H. J. Zhang, S. C. Zhang, X. Dai, and Z. Fang, Science **329**, 61 (2010).

[10] C. X. Liu, X. L. Qi, X. Dai, Z. Fang, and S. C. Zhang, Phys. Rev. Lett. **101**, 146802 (2008).

[11] N. Nagaosa, J. Sinova, S. Onoda, A. H. MacDonald, and N. P. Ong, Rev. Mod. Phys. **82**, 1539 (2010).

[12] K. Nomura and N. Nagaosa, Phys. Rev. Lett. **106**, 166802 (2011).

[13] C. Z. Chang *et al.*, Science **340**, 167 (2013).

[14] J. G. Checkelsky, R. Yoshimi, A. Tsukazaki, K. S. Takahashi, Y. Kozuka, J. Falson, M. Kawasaki, and Y. Tokura, Nat. Phys. **10**, 731 (2014).

[15] X. Kou *et al.*, Phys. Rev. Lett. **113**, 137201 (2014).

[16] X. L. Qi, R. Li, J. Zang, and S. C. Zhang, Science **323**, 1184 (2009).

[17] J. Wang, B. Lian, X. L. Qi, and S. C. Zhang, Phys. Rev. B **92**, 081107 (2015).

[18] T. Morimoto, A. Furusaki, and N. Nagaosa, Phys. Rev. B **92**, 085113 (2015).

[19] J. Wang, Q. Zhou, B. Lian, and S.-C. Zhang, Phys. Rev. B **92**, 064520 (2015).

[20] M. Mogi, M. Kawamura, A. Tsukazaki, R. Yoshimi, K. S. Takahashi, M. Kawasaki, and Y. Tokura, Sci. Adv. **3**, eaao1669 (2017).

[21] M. Mogi, M. Kawamura, R. Yoshimi, A. Tsukazaki, Y. Kozuka, N. Shirakawa, K. S. Takahashi, M. Kawasaki, and Y. Tokura, Nat. Mater. **16**, 516 (2017).

[22] Q. L. He *et al.*, Science **357**, 294 (2017).

[23] D. Xiao *et al.*, Phys. Rev. Lett. **120**, 056801 (2018).

[24] S. Grauer, K. M. Fijalkowski, S. Schreyeck, M. Winnerlein, K. Brunner, R. Thomale, C. Gould, and L. W. Molenkamp, Phys. Rev. Lett. **118**, 246801 (2017).

[25] C. Z. Chang *et al.*, Nat. Mater. **14**, 473 (2015).

[26] X. F. Kou *et al.*, Acs Nano **7**, 9205 (2013).

[27] Y. Ou *et al.*, Adv. Mater. **30**, 1703062 (2018).

[28] C. Liu, Y. Zang, W. Ruan, Y. Gong, K. He, X. Ma, Q. K. Xue, and Y. Wang, Phys. Rev. Lett. **119**, 176809 (2017).

[29] J. S. Lee, A. Richardella, D. W. Rench, R. D. Fraleigh, T. C. Flanagan, J. A. Borchers, J. Tao, and N. Samarth, Phys. Rev. B **89**, 174425 (2014).

[30] J. G. Checkelsky, J. T. Ye, Y. Onose, Y. Iwasa, and Y. Tokura, Nat. Phys. **8**, 729 (2012).

[31] Y. Deng, Y. Yu, M. Z. Shi, Z. Guo, Z. Xu, J. Wang, X. H. Chen, and Y. Zhang, Science **367**, 895 (2020).

[32] C. Liu, Z. Zhang, Y. Ou, X. Ma, K. He, Q. Xue, and Y. Wang, to be published.

[33] F. Wang *et al.*, Nat. Commun. **12**, 79 (2021).

[34] K. Yasuda *et al.*, Nat. Phys. **12**, 555 (2016).

[35] H. Wu *et al.*, Adv. Mater. **32**, e2003380 (2020).

[36] W. Wang *et al.*, Nat. Mater. **18**, 1054 (2019).

[37] Q. L. He *et al.*, Nat. Commun. **9**, 2767 (2018).

[38] C. Liu *et al.*, Nat. Commun. **12**, 4647 (2021).

[39] C. Liu *et al.*, Nat. Mater. **19**, 522 (2020).

[40] Y. S. Hor *et al.*, Phys. Rev. B **81**, 195203 (2010).

[41] D. M. Zhang *et al.*, Phys. Rev. B **86**, 205127 (2012).

[42] C. Z. Chang *et al.*, Adv. Mater. **25**, 1065 (2013).

[43] M. Ye *et al.*, Nat. Commun. **6**, 8913 (2015).

[44] P. Sessi, F. Reis, T. Bathon, K. A. Kokh, O. E. Tereshchenko, and M. Bode, Nat. Commun. **5**, 5349 (2014).

[45] M. D. Li, C. Z. Chang, L. J. Wu, J. Tao, W. W. Zhao, M. H. W. Chan, J. S. Moodera, J. Li, and Y.

M. Zhu, Phys. Rev. Lett. **114**, 146802 (2015).

[46] P. Sessi *et al.*, Nat. Commun. **7**, 12027 (2016).

[47] P. Russmann *et al.*, J Phys-Mater **1**, 015002 (2018).

[48] J. A. Krieger *et al.*, Phys. Rev. B **96**, 184402 (2017).

[49] N. Liu, J. Teng, and Y. Li, Nat. Commun. **9**, 1282 (2018).

[50] Y. Y. Li *et al.*, Adv. Mater. **22**, 4002 (2010).

[51] Y. Zhang *et al.*, Nat. Phys. **6**, 584 (2010).

[52] G. Rosenberg and M. Franz, Phys. Rev. B **85**, 195119 (2012).

[53] Z. Zhang *et al.*, Nat. Commun. **5**, 4915 (2014).

[54] J. S. Dyck, C. Drasar, P. Lost'ak, and C. Uher, Phys. Rev. B **71**, 115214 (2005).

[55] J. S. Dyck, P. Švanda, P. Lošt'ák, J. Horák, W. Chen, and C. Uher, J. Appl. Phys. **94**, 7631 (2003).

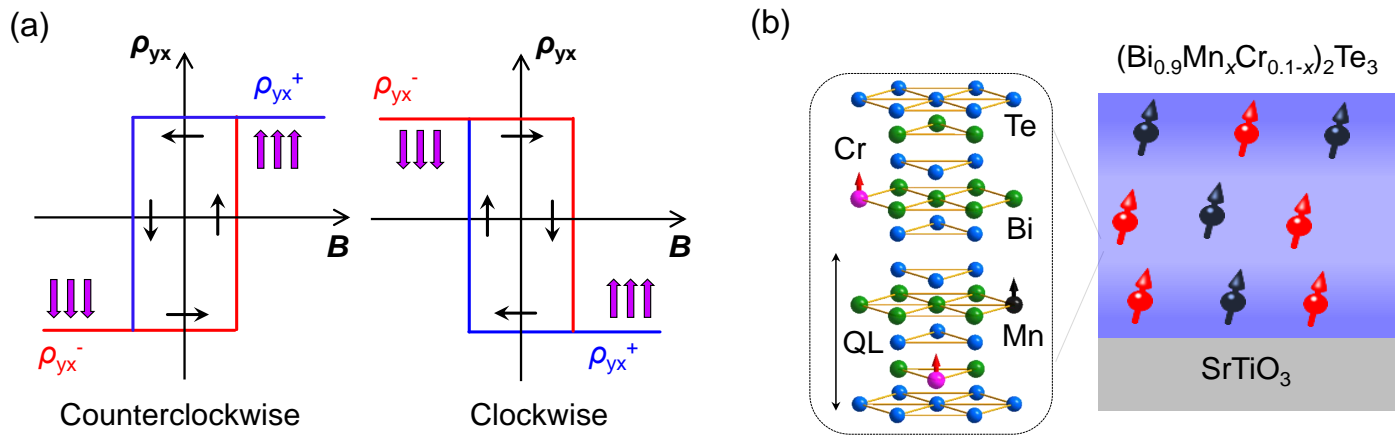

Figure 1

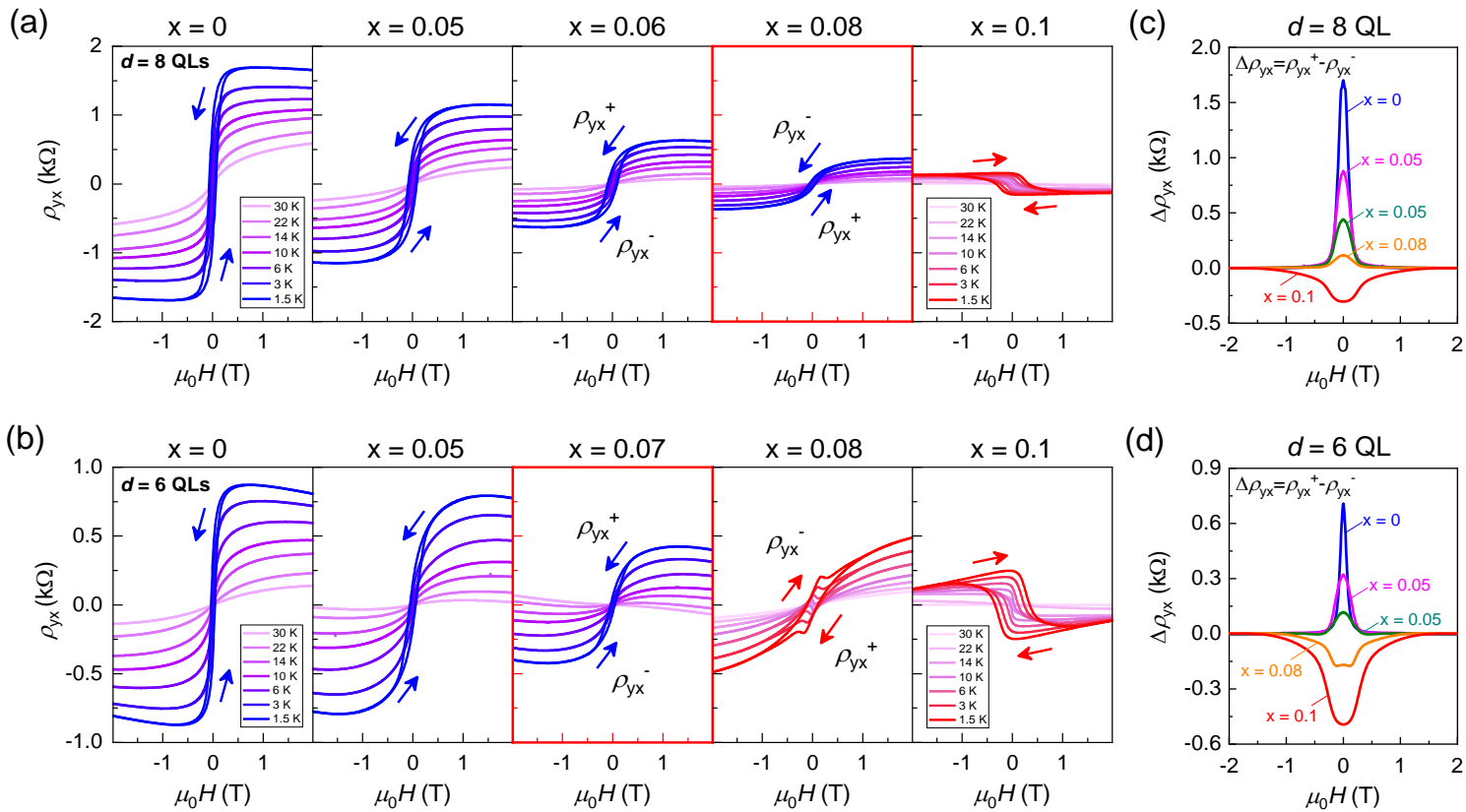

Figure 2

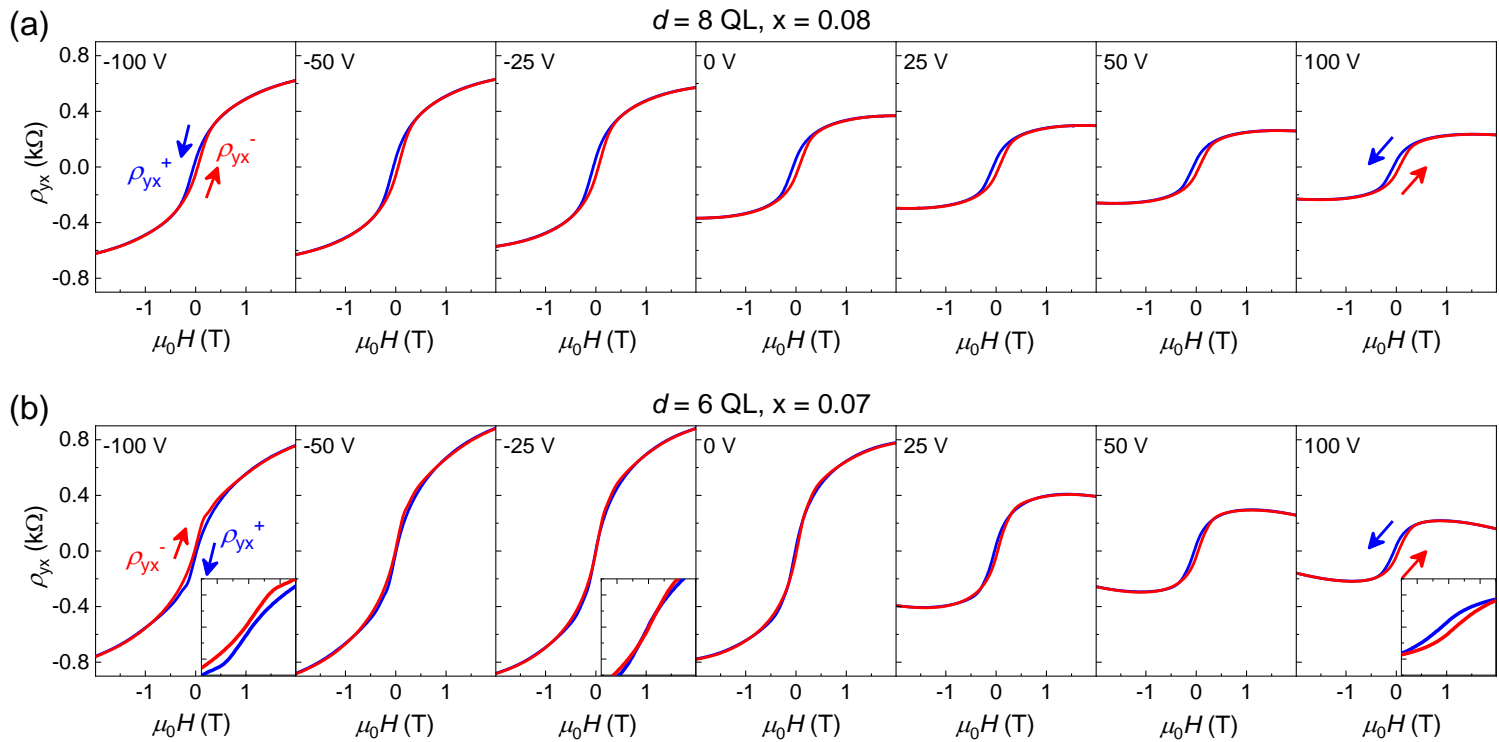

Figure 3

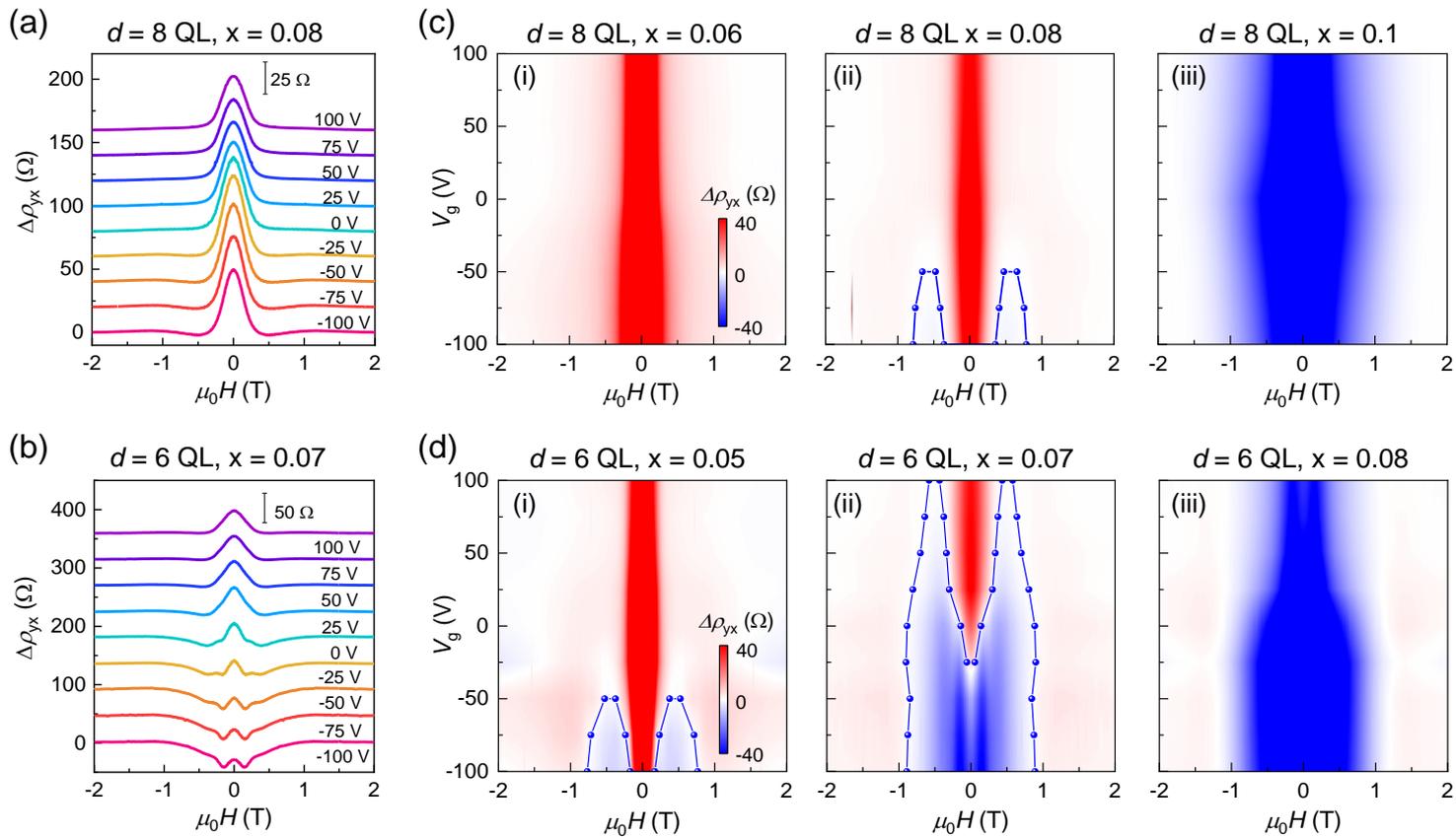

Figure 4

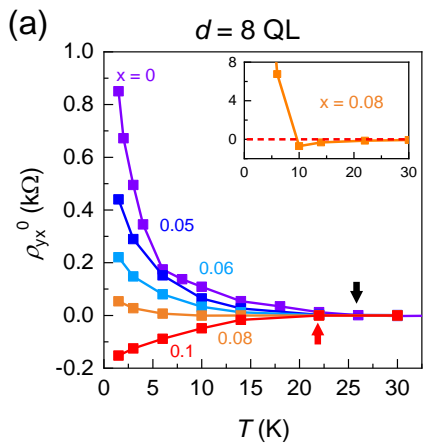
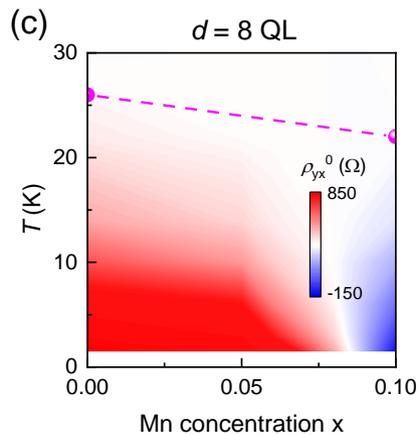
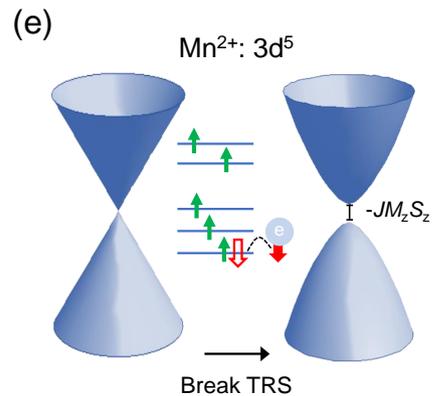
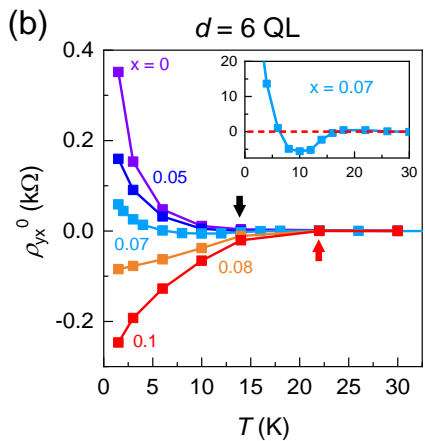
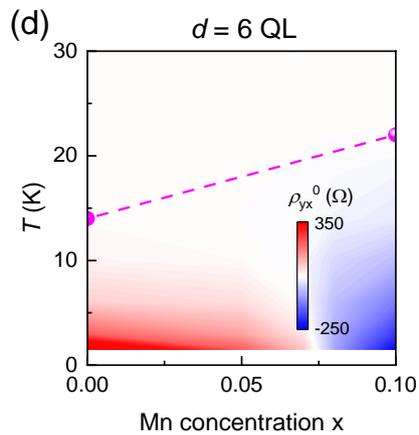
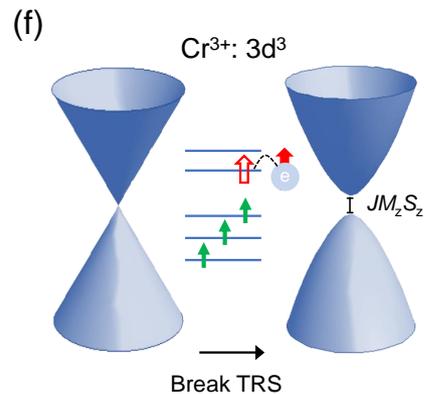

Figure 5